# EMP Lasing from Mn doped lead bromide perovskites nanorods

Shuangyang Zou[#], Zhihong Gong[#], Lipeng Hou, Ruibin Liu, Haizheng Zhong, Bingsuo Zou*

Beijing Key Laboratory of Nanophotonics & Ultrafine Optoelectronic Systems, Beijing Institute of Technology, Beijing 100081, China;

**ABSTRACT:** Diluted magnetic semiconductor (DMS) nanostructures are promising platform to modulate carriers and spins for new information devices. Here we report that the high quality pure $CH_3NH_3PbBr_3$ nanorods and Mn doped $CH_3NH_3PbBr_3$ nanorods have been prepared by solution method and in which the exciton magnetic polarons (EMP) formed in Mn doped NRs, and a single mode lasing phenomenon from collective EMP in single NR have been detected when excited by fs pulse laser. This finding helps to understand the exciton and spin interactions and pave ways to the realization of new type of bosonic laser.

## INTRODUCTION

Organometal halide perovskites, as a novel type of semiconductor, have attracted much attention in the world, they have high optical absorption coefficient, suitable optical band gap, long carrier scattering distance, which could be used for a high quality photoelectric functional material including solar cell[1], photodetector[2], display[3] and lasing[4]. For example, Xin et al succeeded in growing organic–inorganic halide perovskites which have shown tunable lasing behavior.[4] Yang et al have demonstrated all-inorganic perovskites NWs with defined morphology.[5] Zhu et al have demonstrated the lead halide perovskites single-crystal nanowires and its broad stoichiometry-dependent tunable lasing emission.[6] Lasing from films[7] or platelets[8] of hybrid organic–inorganic perovskites has also been reported.

In previous publications on semiconductor material,[4, 5, 7-9] most of lasing behavior come from the photoinduced electron-hole recombination, few of them come directly from the exciton-exciton scattering in nanostructure[10] or exciton-photon coupling cavity. Due to the carrier effect, the laser emissions in the electron-hole recombination system usually have broad band and bad coherence, which limited their applications in the high level applications in the quantum information technology. So the cavity exciton-photon technique to confine high density excitons by photon has been developed.[11] In a recent publication,[12] we proposed a new way to bind multi-excitons by ferromagnetic coupled spins in CdS:Co DMS nanostructure, which can lead to collective EMP to give single mode lasing. Such spin-polarized coherent radiation has important applications in the spin-modulated information technology. So it is important and challenging to find more such ferromagnetic semiconductor materials working for the study on its spin modulation and related applications.

So we have synthesized high quality pure $CH_3NH_3PbBr_3$ nanorods and Mn doped $CH_3NH_3PbBr_3$ nanorods by solution method. The room temperature related photoluminescence(PL) and stimulated emission of $Mn^{2+}$ doped nanorod have been studied, which prove a clear coupling between exciton and spins by EMP formation and lasing, this laser behavior will find applications in many new information technologies.

## EXPERIMENTAL SECTION

$CH_3NH_3Br$ was synthesized by reaction of the methylamine and HBr solution. Firstly, methylamine in absolute ethanol (60mL, 40%) was stirred for 2 hours and cooled to 0 °C with the addition of HBr (65mL, 48%). Then the solution was kept in container without stirring for 1 hour. Then rotary evaporation was applied to evaporate the solvent at 50 °C. The precipitate was washed three times with diethyl ether and dried under vacuum (50 °C, 24 h).

$CH_3NH_3PbBr_3$ nanorods were synthesized by the reaction of $Pb(Ac)_2$ and $CH_3NH_3Br$. Firstly, $Pb(Ac)_2$ film was prepared by spin-coating on glass substrate which was cleaned ultrasonically. Then the $Pb(Ac)_2$ glass substrate was immersed in methylamine in absolute ethanol (1mL, 5mg/mL) for 15 hours. The substrate was dried naturally after cleaned several times by methylamine in absolute ethanol. Finally, $CH_3NH_3PbBr_3$ nanorods formed on the glass substrate.

$Mn^{2+}$ doped $CH_3NH_3PbBr_3$ nanorods were synthesized by the same method. The manganese acetate was added into the $Pb(Ac)_2$ solution before preparing the $Pb(Ac)_2$ film.

**Structural and morphological characterizations.** The morphology and composition of the samples were characterized via scanning electron microscopy (SEM, Zeiss Supra55) equipped with EDS. The phase purity of the product was examined via XRD by using an X-ray diffractometer (Brucker D8-advance) with Cu Kα radiation (1.5406Å), therein maintaining the operating voltage and current at 40 kV and 40 mA, respectively. During the measurement 2θ ranged from 20° to 80° with a step of 0.02° and a count time of 1 s is used.

**Magnetization measurement.** The measurements of the magnetization (M) as a function of H were realized using a superconducting quantum interface device (SQUID, MPMS¬-XL-7, Quantum Design) magnetometer in the range $-7 \leqslant H \leqslant 7T$. The temperature changed from 2K to 300K. All magnetization measurements were performed in the presence of an in-plane magnetic field on the massive sample on the silicon wafer substrate. The measured magnetization data were corrected by accounting for the diamagnetic contribution from the substrate.

**PL and lasing spectrum.** The PL, polarized PL and lasing spectrum were measured by the same laser confocal optical microscopy (Horiba JY iHR550, Olympus BX51 M) using a 405 nm continuous wave (CW) semiconductor laser as an excitation source. For the low-temperature PL measurements, the sample was placed in a microscope cryostat. Liquid nitrogen was used to cool the samples down to 77 K. The lasing spectra of the nanorods were obtained using a femtosecond laser (Coherent, 130fs, 80MHz) for excitation.

RESULTS AND DISCUSSION

The $CH_3NH_3PbBr_3$ NRs were obtained by using solution growth method described in the methods section. The SEM morphology images of as-prepared NRs and Mn doped NRs are shown in Figure 1a, c respectively. Figure 1b is the magnification image of NR. The three images show that the surfaces of NRs are clean and smooth. The length of as-grown NRs is about 50-100μm(?). The elementary composition of the NRs in Figure 1a was C, N, Br and Pb according to the energy dispersive X-ray (EDX) analysis (Figure 1e). The element ratio of Br to Pb is 2.9, quite close to stoichiometry ratio of 3. Figure 1d presents the XRD pattern of these NRs, which confirms their crystal phase and lattice quality. Figure 1f shows the EDS profile of $CH_3NH_3PbBr_3$:Mn NRs, in which the Mn concentration is about 0.21%. In the growth process the manganese concentration tuned from 0.21 to 7.35%(atomic %) can be identified by EDX. According to the XRD pattern, no phase separation of NRs is observed for manganese concentration up to 7.35%.

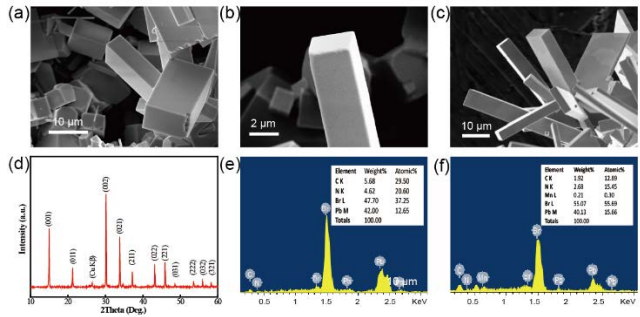

Figure 1. (a) SEM image of as-grown $CH_3NH_3PbBr_3$ NRs. (b) Magnification image of NR. (c) SEM image of as-grown $Mn^{2+}$ doped $CH_3NH_3PbBr_3$ NRs. (d) XRD pattern of the as grown NRs. (e) EDX profile measured from one NR. (f) EDX profile of $Mn^{2+}$ doped $CH_3NH_3PbBr_3$. Inset is the relevant element ratio.

**Magnetic response of Mn-doped $CH_3NH_3PbBr_3$.** Single Mn-doped $CH_3NH_3PbBr_3$ crystal of several μm size could be grown, whose optical image of as-prepared one single-crystal shown in Figure 2a inset. Figure 2a shows their magnetization hysteresis loop line of $CH_3NH_3PbBr_3$:Mn at 0.67% Mn doping as a function of the applied magnetic field (H). The inset is the magnified area near zero magnetic fields. A good magnetic hysteresis loops was shown in M-H curves (-1000~1000Oe), indicating the ferromagnetic behavior of these single-crystal powder. To determine its Curie temperature, we also measured the M-T curve by applying a 7T magnetic field by SQUID, as shown in Figure 2b. The magnetization slowly decreases with increasing temperature, cross point at Curie temperature has not seen even up to room temperature.

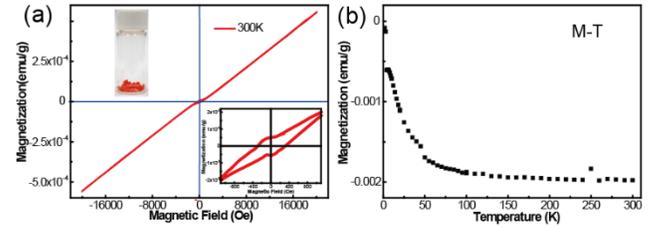

Figure 2. (a) Hysteresis loops (M vs H) of doped $CH_3NH_3PbBr_3$ single-crystal measured at 300K. Inset is the $CH_3NH_3PbBr_3$ single-crystal in the vial. (b) The relationship between Magnetization and Temperature of $CH_3NH_3PbBr_3$ single-crystal.

Figure 3a and Figure 3b shows respective room-temperature PL spectra of pure single $CH_3NH_3PbBr_3$ NR and Mn-doped $CH_3NH_3PbBr_3$ NR excited by 405 nm line of a CW semiconductor laser, its PL image is shown in fig.1S. A very strong band-edge emission band centered at 539 nm was obtained for $CH_3NH_3PbBr_3$ NR (Figure 3a), while two emission peaks at 543nm and 557nm show up for Mn-doped $CH_3NH_3PbBr_3$ NR. Their difference of emission peak lie in two reasons: 1) The band-edge emission peak shows red shift from 539nm to 534nm because of manganese doping, the former energy should occur at bandedge containing absolute components due to the exciton-phonon coupling in a normal polar semiconductor. But in Mn-doped $CH_3NH_3PbBr_3$ NR, the 534nm peak should be the free exciton of $CH_3NH_3PbBr_3$ NR, its enhanced exciton-phonon coupling component should show up in the longer wavelength range than 539nm.[12] 2) The second emission peak at 557nm for $CH_3NH_3PbBr_3$:Mn can be assigned to exciton magnetic polaron (EMP) emission,[13] which is formed due to coupling between ferromagnetic coupled transition metal ion aggregate and photo-induced exciton (FX).[14] Farther from bandedge than FX, so EMP contain some component FX coupling with LO phonon, and is more stable than FX. Because EMP contain spin polarized ions, so its emission have narrow polarized nature, which is not in agreement with the rod axis direction as shown in Figure 3c and Figure 3d, as that Co-doped CdS belt.[12]

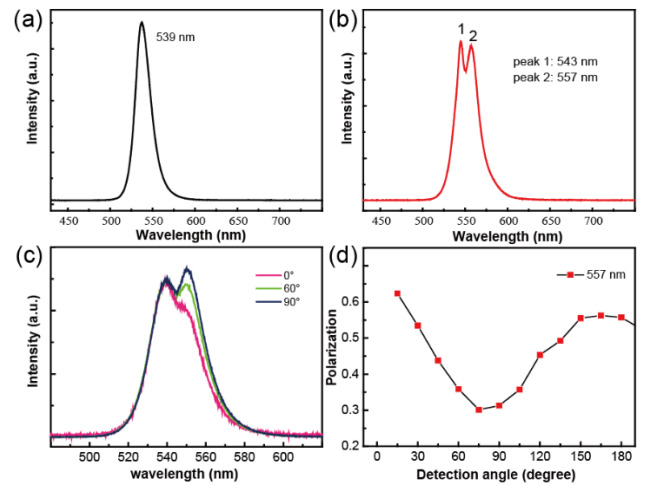

Figure 3. (a) PL emissions from a single pure $CH_3NH_3PbBr_3$ NR and (b) Mn-doped $CH_3NH_3PbBr_3$ NR at room temperature.

(c)The polarized spectra of Mn-doped $CH_3NH_3PbBr_3$ NR. (d) PL intensity ratios $(I_{//}-I\perp)/(I_{//}+I\perp)$ of the Mn-doped NR with emission peak at 557 nm as a function of polarized angles.

Figure 3c shows the polarized PL spectra of Mn-doped $CH_3NH_3PbBr_3$ NRs. The detection angle was changed stepwise from 0° to 90° through a rotating polarizing beam splitter. As shown in Figure 3d, the one NR show strong linearly polarization emission with polarization ration $P=(I_{//}-I\perp)/(I_{//}+I\perp)$ where the $I_{//}$ and $I\perp$ are the intensities parallel $(I_{//})$ and perpendicular $(I\perp)$ to the long axis of the NRs, which show no agreement between them. This difference indicates that the spin polarization has not relationship to the rod axis.

Figure 4a shows the PL spectra of $CH_3NH_3PbBr_3$ with 0.21% Mn(II) doping concentration excited by a fs laser (400 nm) of varied powers, whose image at low power is shown in fig.2S. PL spectra from spontaneous emission to stimulated emission at 555.5nm can be seen clearly. The PL emission intensity and peak FWHM is extracted and plotted as a function of pumping power (Figure 4b). When the excitation intensity was below the threshold (0.35 $\mu j/cm^2$), a broad spontaneous emission band was observed. However, the emission intensity shows a superlinear rising with excitation power over the threshold. Moreover, the peak FWHM becomes narrow (even to 1.45nm width) with rising pumping power. The most important phenomenon here deserve to be emphasized is its single mode lasing, which is much different from that the multi-mode lasing due to cavity effect in nanowire.[15] The luminescence schematic of FX and EMP is shown in Figure 4d.

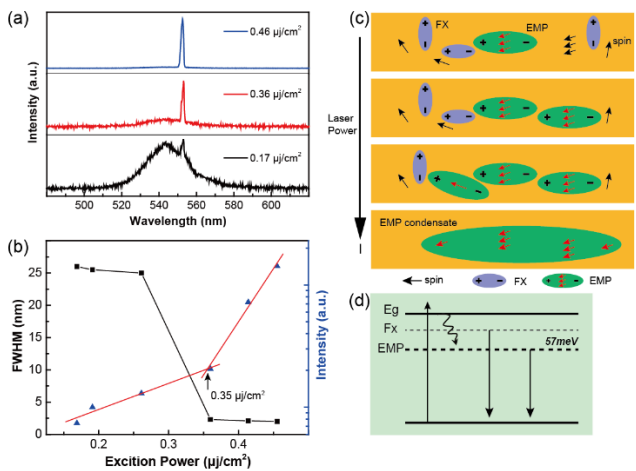

Figure 4. (a) The simulated emission of $CH_3NH_3PbBr_3$ NRs with 0.21% Mn(II) doping concentrations. (b) The PL emission intensity and FWHM is extracted and plotted as a function of pumping power. (c) The schematic of EMPs formation in perovskite. (d) The luminescence schematic of FX and EMP.

**Discussion:** The ferromagnetic mechanism of DMS usually comes from the spin-spin coupling and spin-carrier interactions. The former usually work on the wide-gap semiconductor with a high Curie temperature, while the latter dominate in the narrow gap semiconductor with a low Curie temperature.[16] The $CH_3NH_3PbBr_3$ should be the wide-gap semiconductor, so its magnetization as measured in Figure 2a is mainly due to Mn-Mn ferromagnetic coupling or $(MnBr)_n$ cluster as shown in CdS:Mn.[14] Li and coworkers just proposed its potential application in optical spintronics.[17] Here our experiments indicate another spin-related behavior as described in next section.

After photoexcitation above bandgap, the produced exciton could couple with ferromagnetic coupled cluster to form EMP. Such EMP has a much larger binding energy than FX. From separation of FX and EMP emission peaks in Figure 3b, the binding energy difference for EMP to FX is 57meV. FXs in $CH_3NH_3PbBr_3$ often easily dissociate into electron and hole to produce electron hole separation, hence to be used for solar cell.[1] But EMP cannot dissociate at room temperature due to larger binding energy. Moreover, EMP is the coupling composite of FX, coupled spins and LO phonon, all with bosonic nature. This bosonic composite excitation in DMS nanostructure can produce collective EMP,[18] like a nonequilibrium bosonic condensate, to coherently emit light.[12] This type of lasing seems to look like that exciton-polariton laser, the exciton-photon coupling excitation, in a semiconductor microcavity,[11] in which the photon by an outer laser is used to confine exciton within a microcavity, then produce bosonic lasing. However, the collective EMP lasing need not a cavity, the confining force comes from the intrinsic ferromagnetic spin cluster inside the DMS nanostructure. This process can be described in Figure 4c. All above optical experimental results were obtained at room temperature. Therefore a ferromagnetic spin cluster formation in a DMS nanostructure is a crucial technique to tune the exciton and EMP coupling, hence obtaining their EMP luminescence and lasing at room temperature. This DMS nanostructure may be a typical platform for the spin modulated exciton system.

The $CH_3NH_3PbBr_3$:Mn QDs have also been obtained with tunable emission and Mn ion emission band,[19] we have also studied them, but have not detected their clear ferromagnetic properties and related EMP response, so EMP effect as above should happen in large size at room temperature, which need further verification.

CONCLUSION

High quality $CH_3NH_3PbBr_3$, $CH_3NH_3PbBr_3$:Mn NRs and their single-crystals have been synthesized. The Mn(II) doping $CH_3NH_3PbBr_3$ behaves with DMS nature. The PL emission and magnetic behavior of NRs in a range of Mn doping concentration indicate a new excitation formation---EMP with higher stability than FX. This EMP come from the interactions between exciton and ferromagnetic coupled TM ions. High density EMP can form collective EMP to produce single mode lasing emission with specific polarization profile, which is proved not depend on the formation of optical cavity. Therefore the Mn doped $CH_3NH_3PbBr_3$ perovskite can be used to study the exciton condensate behavior and find applications in the spin-polarized photonics devices and quantum information processing in the future.

AUTHOR INFORMATION

**Corresponding Author**

* E-mail: zoubs@bit.edu.cn.   phone: 86-10-68910204

**Author Contributions**

The manuscript was written through contributions of all authors.

**Funding Sources**

The authors thank the 973 project (2014CB920903) of MOST for financial support.

**Notes**

The authors declare no competing financial interests.

## ACKNOWLEDGMENT

(Word Style "TD_Acknowledgments"). Generally the last paragraph of the paper is the place to acknowledge people (dedications), places, and financing (you may state grant numbers and sponsors here). Follow the journal's guidelines on what to include in the Acknowledgement section.

## ABBREVIATIONS

EMP, exciton magnetic polaron; LEMPs, localized exciton magnetic polaron; PL, photoluminescence; SEM, scanning electron microscopy; EDX, energy dispersive X-ray; XRD, X-ray diffractometer; SQUID, superconducting quantum interface device; CW, continuous wave; H, magnetic field; M, magnetization; NRs, nanorods; FWHM, the width at half maximum;